\documentclass[nofootinbib,twocolumn,superscriptaddress,letterpaper]{revtex4}
\usepackage{graphicx}
\usepackage{amssymb}
\usepackage{amsmath}

\def\mysection#1{{\bf #1.} }

\def\lsim{\mathrel{\rlap{\lower4pt\hbox{\hskip1pt$\sim$}}
    \raise1pt\hbox{$<$}}}         
\def\gsim{\mathrel{\rlap{\lower4pt\hbox{\hskip1pt$\sim$}}
    \raise1pt\hbox{$>$}}}         
\def\ttbar{t\bar t}
\def\qqbar{q\bar q}
\def\mttbar{m_{t\bar t}}
\def\PLR{P_{LR}}
\def\mbl{m_{bl}}
\def\mkkg{M_{KKG}}

\begin{document}

\begin{titlepage}
 
\hfill$\vcenter{
\hbox{\small BNL-HET-06/13,
      MSUHEP-060915, SU-4252-838, YITP-SB-06-43} }$ 

\begin{center}
{\Large \bf LHC Signals from Warped Extra Dimensions}
\vskip .1in
{\bf Kaustubh Agashe}$^1$,  
{\bf Alexander Belyaev}$^2$,
{\bf  Tadas Krupovnickas}$^3$,
{\bf Gilad Perez}$^4$
and {\bf Joseph Virzi}$^5$\\  
{\em $^1$ Department of Physics,
Syracuse University, 
Syracuse, NY 13244}\\ 
{\em $^2$ 
Department of Physics and Astronomy, 
Michigan State University, East Lansing, MI 48824} \\ 
{\em $^3$ 
Brookhaven National Laboratory, 
Upton, NY 11973} \\ 
 {\em $^4$
C.N. Yang Institute for Theoretical Physics,
         State University of New York,
         Stony Brook, NY 11794-3840}\\  
{\em $^5$ Lawrence Berkeley National Laboratory,
Physics Division, 
1 Cyclotron Road,
Berkeley, CA 94720}

\end{center}

\begin{center} {\bf Abstract}\\\end{center}

We study production of Kaluza-Klein gluons (KKG) at the Large Hadron Collider (LHC)
in the framework of a warped extra dimension with the Standard Model (SM) fields propagating in the bulk.
We show that the detection of KK gluon is challenging since
its production is suppressed by small couplings to the proton's constituents.
Moreover, the KK gluon decays mostly to top pairs due to an enhanced coupling and hence is broad.
Nevertheless, we demonstrate that for $M_{KKG} \lesssim$ 4 TeV, 100 fb$^{-1}$ of data at the LHC can provide discovery of the KK gluon. 
We utilize a sizeable left-right polarization asymmetry from the KK gluon resonance to maximize the signal significance, 
and we explore the novel feature of extremely highly energetic ``top-jets''.
We briefly discuss how the detection of electroweak gauge KK states ($Z/W$) faces a similar challenge 
since their leptonic decays (``golden'' modes) are suppressed.
Our analysis suggests that other frameworks, for example little Higgs, 
which rely on UV completion via strong dynamics might face similar challenges, namely
(1) Suppressed production rates for the new particles (such as $Z^{ \prime}$), due to their ``light-fermion-phobic'' nature, and
(2) Difficulties in detection since the new particles are broad and decay predominantly to third generation quarks and longitudinal gauge bosons.
\vskip .2in

\end{titlepage}
\newpage
\renewcommand{\thepage}{\arabic{page}}
\setcounter{page}{1}

\section{Introduction}
Solutions to the Planck-weak hierarchy problem of the SM typically
invoke new particles charged under the SM at the TeV scale. The lore is
that such particles will be readily accessible at the LHC, especially
the strongly interacting ones. In this paper, we consider the solution
to the hierarchy problem based on the Randall-Sundrum I (RS1) framework of a warped extra dimension
\cite{rs1}.
Specifically, we consider this framework with the SM gauge and fermion
fields propagating
in the bulk of the warped extra dimension,
which provides a solution to the flavor puzzle of the SM as well.
We focus on detecting the Kaluza-Klein (KK) partner of the SM gluon
at the LHC -- as we explain, KK gluon is probably the best channel
to probe the  RS1 framework. We show that, despite it being strongly
interacting, it is quite challenging to see a signal from this
particle with a mass of several TeV at the LHC.  The reason is
related to the special (but well-motivated) nature of its couplings
which are non-universal and are ``proton-phobic''. The consequence
of such couplings is that our signal (an excess of top pairs) is
comparable in size to the SM background. With the techniques
developed herein, it should be possible to extract a signal for the
KK gluon (in this framework) at the LHC with $\simeq$ 100 fb$^{-1}$
of data.

The framework involves a slice of AdS$_5$~\cite{rs1}.  Due to the warped
geometry,
the relationship between the $5D$ mass scales (taken to be of order the $4D$
Planck scale) and those in an effective $4D$ description depends on the
location
in the extra dimension. The $4D$ (or zero-mode) graviton is localized near
the
``UV/Planck'' brane which has a Planckian fundamental scale,  whereas the
Higgs
sector is localized near the ``IR/TeV'' brane where it is protected by a
warped-down fundamental scale of order TeV.  This large hierarchy of
scales can be
generated via a modest-size radius of the extra dimension.  Furthermore,
based on
the AdS/CFT correspondence \cite{Maldacena:1997re},  the RS1 model is
conjectured
to be dual to $4D$ composite Higgs models \cite{Arkani-Hamed:2000ds}. %
Hence, our
results might apply in general to $4D$ models with TeV-scale strong dynamics
driving electroweak symmetry breaking (EWSB).

In the RS1 model, the entire SM (including the fermions and gauge bosons) is
assumed to be localized on the TeV brane. The higher-dimensional operators
in the
$5D$ effective field theory (from cut-off physics) are suppressed only by the
warped-down scale $\sim$ TeV,  giving too large contributions to FCNC
processes
and observables related to SM electroweak precision tests (EWPT).
Moreover, this
set-up provides no understanding of the flavor puzzle.

An attractive solution to this problem is to allow the SM fields to
propagate in
the extra dimension \cite{bulkgauge, gn, gp}. In this scenario, the SM
particles
are identified with the zero-modes of the $5D$ fields and the profile of a SM
fermion in the extra dimension depends on its $5D$ mass parameter. We can
then
choose to localize 1st and 2nd generation fermions near the Planck brane
so that
the FCNC's from higher-dimensional operators are suppressed by scales
$\gg$ TeV
which is the cut-off at the location of these fermions~\cite{gp, hs}.
Similarly,
contributions to EWPT are also suppressed.

As a bonus, we obtain a solution to the flavor puzzle in the sense that
hierarchies
in the SM Yukawa couplings arise without introducing hierarchies in the
fundamental
$5D$ theory~\cite{gn, gp, hs}. The 1st/2nd generation fermions have small
Yukawa
couplings to Higgs, which is localized near the TeV brane.  Similarly, the
top quark
can be localized near the TeV brane to account for its large Yukawa coupling.

In this scenario, there are new contributions to EWPT and FCNC's
calculable in the
$5D$ effective field theory (EFT) from KK modes. In particular, the
couplings of SM
fermions to gauge KK modes are non-universal due to the different profiles
for the
SM fermions, resulting in FCNC's. However, the gauge KK modes are
localized near the
TeV brane while the light fermions are near the Planck brane and hence  it
can be
shown that the non-universal part of these couplings are proportional to
the SM
Yukawa couplings~\cite{gp, hs}. Thus, most of the couplings to the new
degrees of
freedom are small and hierarchical, leading to the same symmetry structure
which
suppresses the SM flavor-violating contributions~\cite{aps}  (for recent
related
discussions and the experimental status see~\cite{NMFV}).   The gauge KK
modes also
give contributions to EWPT.  The constraints from the oblique ($S$ and $T$)
parameters can be satisfied with a KK mass scale as low as $\sim 3$ TeV  if a
custodial isospin symmetry is incorporated~\cite{custodial}.

Let us examine the top/bottom sector in detail since the associated
couplings will be relevant for the signals. It is clear that both
$t_{L,R}$ being near the Planck brane gives too small a top Yukawa
coupling. On the other hand, the fact that $(t,b)_L$ is close to the TeV
brane leads to its large coupling to KK $Z$ and, in turn,  results in a
non-universal shift in its coupling to the SM $Z$ via mixing of KK $Z$
with zero-mode $Z$~\cite{custodial}:
$
\delta g_Z^{ b_L }  \sim  g_{Z^{\rm KK }}^{ b_L }
\xi
\frac{ m^2_Z }{ M_{ KK Z }^2 }$
where $\xi\equiv\sqrt{ \log \left( M_{ Pl } / \hbox{ TeV } \right) }$
and $g_{Z^{\rm KK }}^{ b_L }$ is the corresponding non-universal KK $Z$
coupling.
The constraint from data is that $\delta g_Z^{ b_L } / g_Z \lsim 1/4 \%$.

Thus, for a KK scale $\simeq$ a few TeV, there is a tension between obtaining
large top mass and EWPT (i.e., $Z \bar b_L b_L$ coupling) which can be
relaxed
by the following setup: (i) $(t,b)_L$ quasi-localized near TeV brane so
that the
shift in coupling of $b_L$ to $Z$ is on the edge, (ii) $t_R$ localized very
close to TeV brane to obtain large top quark mass and (iii) largest
dimensionless $5D$ Yukawa consistent with perturbativity. Note that the
resulting coupling of $b_L$ to gauge KK modes (including gluon) is
comparable to
the SM couplings and thus is still larger than what is expected on the
basis of
$m_b$ alone, since it is dictated by the large top mass instead. Even with
these
choices, the KK scale is required to be rather high, $\lesssim~5$~TeV.  In
this
case, the couplings of $t_R$, which is localized very near the TeV brane,
to the
gauge KK modes are enhanced: $g^{ t_R }_{ \hbox{SM}^{ \rm KK } }  \sim  g_{
\hbox{SM} }  \xi \,.$

However, such corrections to $Z \bar{b}_L b_L$ coupling can be suppressed by
suitable choice of representation of top and bottom quarks under the
custodial
isospin symmetry \cite{Zbb}. In this case, we can have the other extreme
situation:
$(t,b)_L$ can be localized very close to the TeV brane with $t_R$ being
close to
flat. Also, there is an intermediate possibility with both $(t,b)_L$ and
$t_R$ being
localized close (but not too close) to the TeV brane. The KK scale can
then be as
low as $\sim 3$ TeV for certain choice of profiles for $t_R$ and $(t,b)_L$
 in the
extra dimension \cite{Carena:2006bn}.

In this paper we will consider models with the assignment of reference
\cite{custodial} for the quantum numbers of top and bottom quarks. Based
on the
above profiles, it can be shown that the couplings of KK gluon (and in
general all
gauge KK modes) to light fermions (including $b_R$) are suppressed by 
$\xi$ with
respect to the SM gauge couplings. The coupling to $t_L,b_L$ is neither
suppressed
nor enhanced and only the coupling to $t_R$  (which is practically on the
TeV brane
or composite in the dual $4D$ picture) is enhanced by $\xi$. It can also
be shown
that there is no coupling of single KK gauge field to two  SM gauge bosons at
leading order due to orthonormality of profiles of these particles. To
summarize
(see for example~\cite{aps} for more details) the relevant coupling to the
KK gauge
states can be described,  neglecting effects related to EWSB, via ratio of
RS1-to-SM
gauge coupling

\begin{eqnarray}
{g_{\rm RS}^{q\bar q,l\bar l\, G^{1}}\over g_{\rm SM}}&\simeq&
\xi^{-1}\approx {1\over5}\,, \, \, \, \,
{g_{\rm RS}^{Q3\bar Q3 G^{1}}\over g_{\rm
    SM}} \approx 1\,, \nonumber\\
{g_{\rm RS}^{t_R\bar t_R G^{1}}\over g_{\rm
    SM}} &\simeq& \xi \approx 5 \,, \,\, \, \,
  {g_{\rm RS}^{ GG G^{1}}\over g_{\rm
    SM}}\approx 0 \,, \label{couplings}
\end{eqnarray}
where $q=u,d,s,c,b_R$, $l =$ leptons, $Q^3= (t, b)_L$, $G,G^1$ correspond
to SM and first KK states of the gauge fields respectively and
$g_{\rm RS}^{xyz}, g_{\rm SM}$ stands for the RS1 and the three SM (i.e.,
$4D$) gauge couplings respectively.

It is straightforward to modify our analysis as to accomodate generic
couplings of
the KK gauge fields to the SM third generation quarks.  This will cover the
signals of models with custodial symmetry for $Zb\bar b$~\cite{Zbb}.
However, we
choose to show the explicit results within one scenario to make the steps
of our
analysis and our results more transparent.  A brief discussion of the
signals in
the case where the custodial symmetry for $Zb\bar b$~\cite{Zbb} is
realized is
given in section~\ref{SecZbb}.

We will mostly focus on LHC signals from KK gluons which have the largest
production rate.
The KK mass scale is assume to be $\simeq$ a few TeV.
In cases where a specific mass was required for our analysis a 3$\,$TeV
mass was used.
We also briefly discuss other interesting signals related to the
electroweak gauge KK sector whose detection might be more challenging than
KK gluon,
partly due to lower production rates than for KK gluons and also due to
suppression of decays to ``golden'' modes such as leptons.
In general, the EW sector is also more model dependent.
Earlier studies of KK gluon production at the
LHC~\cite{KKgttbar, Davoudiasl:2000wi} did not consider the effect of the
fermion profiles
which now is understood to be mandatory for the phenomenological viability
of the framework.


\section{LHC signals}

The primary challenge in obtaining a signal at the LHC for gauge KK modes is that the  production is suppressed 
due to the small couplings to the proton constituents as seen in Eq.~\ref{couplings}.

We used both CalcHEP 2.42~\cite{Pukhov:2004ca} and Sherpa version 
1.0.8~\cite{Sherpa}
\footnote{The authors are grateful to the Sherpa team, especially Tanju Gleisberg, for the help in embedding the RS1 KK gluon into Sherpa.}
for the numerical calculations.
The CTEQ6M parton distribution function (PDF) with the QCD renormalization and factorization scales equal to the KK gluon mass ($\mkkg$) was used in CalcHEP 2.42.
The CTEQ6L1 PDF set was used in Sherpa, employing a running scheme for $\alpha_S$ with $\alpha_S(M_Z)=0.118$.
We find that the results do not change significantly between the two 
PDF sets
\footnote{This should not be interpreted as indication of small uncertainties due to PDF's in the cross section since 
the two PDF sets might be correlated.
One of the main points of our study is to identify observables 
which depend rather weakly on the PDF's uncertainties.}. 

\begin{figure}[htbp]
\includegraphics[width=0.48\textwidth]{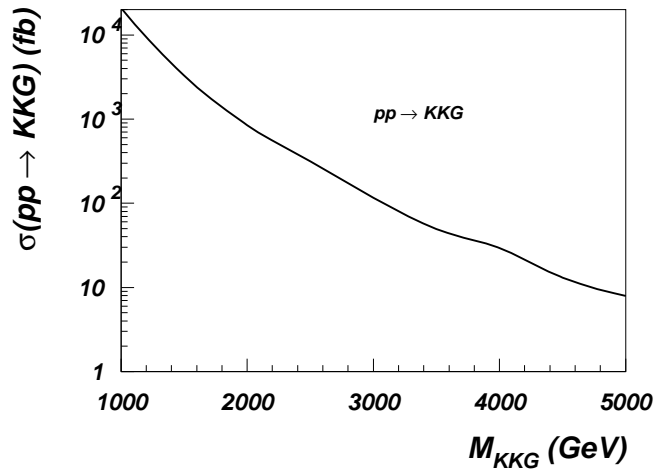}
\caption{\label{kkg_1_signp}
The total cross section of KK gluon production at the LHC as a
function of its mass ($M_{KKG}$).
}
\end{figure}

For KK gluons, CalcHEP yields a moderate cross-section of $\sim 100\,$fb for $\mkkg \sim 3\,$TeV as indicated in Fig.~\ref{kkg_1_signp}.
The cross section falls very quickly for higher KK masses, where for $\mkkg \sim 5$ TeV the cross-section drops to $\sim 10$ fb - 
probably beyond the reach of LHC (as discussed below). 
The dominant production mechanism is through $u\bar{u},d\bar{d}$ annihilation.
We note the production rate for the EW KK gauge fields is suppressed by $(g_Z/g_{ QCD })^2$  relative to KK gluon production.

\begin{figure}[htbp]
\includegraphics[width=0.48\textwidth]{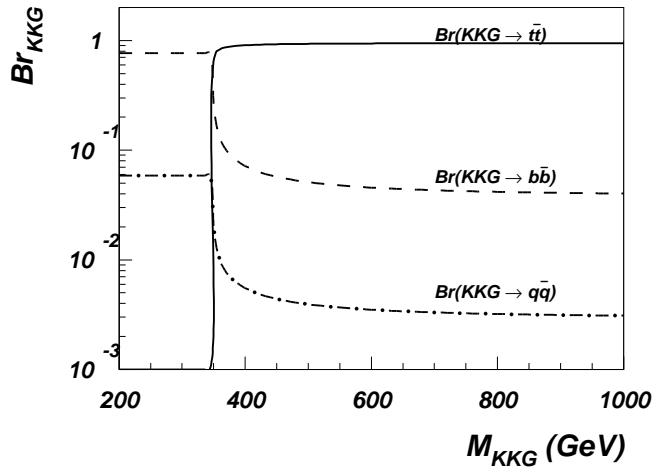}
\caption{\label{kkg_3_br}
The branching ratios of the  KK gluon as a function of its mass.
}
\end{figure}

Another challenge is that, based again on the couplings in Eq.~\ref{couplings},
the fermionic decays of the gauge KK particles (in general) are expected to be dominated by the 3rd generation quarks, 
especially the top quark, due to enhancement of the corresponding couplings. 
For example, the branching ratios for KK gluon decay are shown in Fig.~\ref{kkg_3_br}.
In the case of EW gauge KK modes ($W/Z$), decays to longitudinal weak gauge bosons and the Higgs field are also important due to similarly enhanced couplings.
In particular, the leptonic decay channel for KK $Z$ is highly suppressed. 
In the absence of golden decays modes for KK $Z/W$, we focus 
on signals for the KK gluon which has the larger production 
cross-section.
\footnote{For a related work on KK gluon but with universal couplings 
see~\cite{KKgttbar, Davoudiasl:2000wi}.}

\begin{figure}[htbp]
\includegraphics[width=0.48\textwidth]{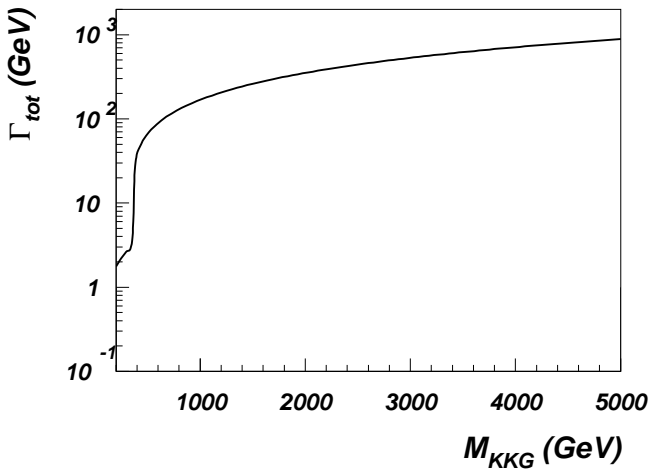}
\caption{\label{kkg_2_gtot}
The total decay width of KK gluon  as a function of its mass}
\end{figure}

A third challenge is related to the fact that due to the strong coupling to top pair (and in case of KK $W/Z$ to Higgs and longitudinal $W/Z$),
a heavy gauge KK mode is rather broad. 
For example, a KK gluon above 1 TeV (as required by precision tests) has decay width of about $M_{KKG}/6$ as presented in
Fig.~\ref{kkg_2_gtot}.
Decay widths of KK $Z/W$ are smaller by $\sim (g_Z/g_{ QCD })^2$.
This large width of KK gauge states creates additional problems for discriminating signal against the background.

\subsection{KK gluons}

In the interesting region of $M_{ KK G }$, well above the $t\bar{t}$ threshold,
the KK gluon decays mainly to $t\bar{t}$ with the branching ratio of about $95\%$ (see Fig.~\ref{kkg_3_br}). 
Hence,
our main focus here will be on the (ultra-relativistic) 
$t\bar{t}$ pairs from decays of KK gluons.
\footnote{For the decays of KK gluon to light quarks (which has small BR in any case), 
the SM QCD background will also be very large.}
Within the SM there are two dominant production mechanisms for
$t \bar{t}$, namely $gg$ (gluon fusion) and $q \bar q$ (quark pair annihilation).
At the LHC, $\ttbar$ production proceeds primarily through gluon fusion~\cite{topLHC}.
\footnote{In the region of interest here, i.e., $\mttbar^2 \approx ( 3 \; \hbox{TeV} )^2$, 
the rate for gluon fusion into top pairs in SM is roughly $4$ times larger than the $q \bar{q}$ annihilation rate.}
$\ttbar$ (top pair) production near threshold has been extensively studied 
(see {\it e.g.} \cite{ttbar} and references therein).
Away from threshold, this simple picture is modified due to the presence of states of higher angular momentum. 
In the other extreme, ultra-relativistic case 
($\mttbar^2\gg 4 m_t^2$) which is the focus of this paper, 
another rather simple and very interesting description emerges~\cite{KPV}.
We make use of the fact that in this limit the SM effects related to EWSB are small and also the top quark chirality is conserved 
(the relevant issues are discussed below when the polarization asymmetry is studied).

The crucial point is that we find, unlike the case in previous 
studies~\cite{KKgttbar, Davoudiasl:2000wi},
the cross-section for SM $t \bar{t}$ production
(in the region $\mttbar - \mkkg \sim \pm \Gamma$)
is {\em comparable} to $t \bar{t}$ production from KK gluons.
Moreover, the SM cross-section has a large uncertainty from gluon PDF's in the large $x$-region~\cite{Pumplin:2002vw}.
Hence, even with $\mkkg$ lighter than 5 TeV obtaining a clear and robust signal is a non-trivial task.
In particular, a simple ``number-counting'' experiment is not enough.
We follow a multi-step strategy to get clear and significant results.
We first consider the differential top pair cross-section. 
Then we analyse a left-right polarization asymmetry, expected to have a clean and robust prediction
for ultra-relativistic top quarks~\cite{KPV} in the SM and our framework.
The combination of the two observables yields a powerful tool to probe our class of models.

\subsubsection{Event Generation and Jet Reconstruction}
Sherpa version 1.0.8, using a customized class to implement the appropriate vertices, was used to generate events, using LHC parameters. 
A cone jet algorithm with $\Delta R=0.4$~\cite{cone_jet_algorithm}, or
C4 for short, was used to reconstruct jets ($\Delta R=\sqrt{\Delta \eta^2+\Delta\phi^2}$). 
Events were generated with cross sections calculated to leading order. 
We do not analyze the effects of pile-up, 
nor characterize the underlying event.
In addition, we do not include detector effects.

\subsubsection{Details of analysis}
In this section, we discuss in more detail how we performed our analysis.
Our preferred reconstruction mode is $t\bar{t} \rightarrow b\bar{b}jjl\nu$ (semileptonic), whose signature we refer to simply as ``$lepton+jet$'' ($lj$).
We use the terms hadronic- and leptonic tops to refer to those quarks which decay into the hadronic mode and leptonic modes, respectively.
We focus primarily on the SM irreducible background from $\ttbar$ production 
and discuss several crucial aspects of the dominant reducible
background,
$W+$jets and single top production.

For the leptonic side reconstruction,
we searched for high $P_T$ leptons, presumably excluded from jets.
We will refer to this condition as isolation, 
and we will discuss this point in more detail below. 
We assumed that the $W$ from the decay of a top quark further decayed
leptonically,
inferring an (undetected) neutrino to 
account for the missing transverse energy. 
A $b$-jet was required to combine with the $W$ to form an on-mass-shell top quark, via an invariant mass condition ($m_{Wb} = M_{t}^{lep} = M_{t} \pm 50\,$GeV).

We now develop the methods of hadronic side reconstruction, but we must first place them in context.
The extremely energetic nature of the top quarks in our signal ($P_T >
1\,$TeV) leads us to deviate from the hadronic top reconstruction
methods (see {\it e.g.}~\cite{ttbarspin}), 
where they studied $\ttbar$ production with $\mttbar \lesssim 600\,$GeV.
\footnote{The energy regime $P_{T} \gsim 600\ GeV$ for jet reconstruction has not been extensively studied.}
Top quarks with $P_T > 1.0\,$ TeV tend to produce highly collimated jets. 
We focused on the C4 algorithm, which will not resolve higher jet multiplicities in high $P_{T}\ \ttbar$ events.
Reducing the cone size to $R=0.2$, for example, only masks this issue, and we eventually succumb to the same problem. 
This renders the hadronic top quark ($t \rightarrow bjj$) reconstruction mode in~\cite{ttbarspin} far too inefficient for our purposes.
Note also that the $\Delta R$ lepton to $b$-jet isolation criterion
(from the leptonic top) falls into this trap for the same reasons.
We propose a different strategy as follows:

(1) In searching for an isolated lepton,  we modify the $(\Delta R)$ leptonic top
reconstruction mode (see {\it e.g.}~\cite{ttbarspin}),
augmenting the lepton to $b$-jet isolation criteria with an
energy scale-invariant cut. The lepton is considered isolated from a given jet
(light jet, $b$-jet, etc.) if they are separated by an angular distance $\Delta R >
0.4$. If a lepton is found inside the cone of a $b$-jet, the lepton is removed from
the $b$-jet and the $b$-jet is reclustered, in which case the invariant mass of the
lepton and $b$-jet system must satisfy $\mbl>40\,$GeV. $\mbl$ provides a measure of
the relative transverse momentum between the lepton and the $b$-jet. 
So, for $b$-quark and lepton isolation
we apply $\Delta R >0.4$ {\it or} $\mbl>40\,$GeV cut,
while
$\Delta R=0.4$ isolation criterion between lepton and all other jets remains in effect.

(2) If the jet multiplicity allows (2 $b$-jets and $\ge2$ light jets), 
we require that the invariant mass of the light jets reconstruct a $W$
according to parameters in
Table~\ref{tab:sel_cuts}.
The invariant mass, $M_{t}^{had}$ of the ($W$ + hadronic $b$-jet)-system is required to reconstruct a top quark according to parameters is  Table~\ref{tab:sel_cuts}.

In a dijet event with 1 $b$-jet, if the other jet has $P_{T} > 800\,,$GeV, we tag it as 
``top (or $t$)-jet'', 
keeping only $\phi$, $\eta$ and transverse energy information from the reconstruction.
\footnote{In principle, a more sophisticated analysis would consider substructure resolution within this jet. 
The authors thank Frank Paige for discussions on this issue.}
We {\em impose} a top-mass hypothesis on the jet, setting $M_{jet} \rightarrow M_{top} = 174.3\,$GeV.
\begin{center}
$P_{x} = P_{T} \cos ( \phi )$ \\
$P_{y} = P_{T} \sin ( \phi )$ \\
$P_{z} = P_{T} \sinh ( \eta )$ \\
$M = \sqrt{E^2-P^2} \rightarrow 174.3\,$GeV
\end{center}

The $t$-jet reconstructed mode dominates the reconstructed signal for
$\mttbar \gsim 2\,$ TeV ($\mttbar$ stands for the top pair invariant mass).
We recuperated a large sample of signal events that we otherwise would have lost via more conventional reconstruction methods.
It would appear the top jet approach would introduce a large
background from such processes as $W$+jets
($Wjj$) and single top production,
especially since we relax the $b$-jet tagging on the hadronic side.
The $P_{T}$ cut is crucial in reducing this background to almost negligible levels. 
We examined the effect of the background by simulating the largest possible sources $Wjj$ (the dominant background), using both CalcHep and Sherpa.
We found that the cross section that satisfies our preselection cuts, 
$m_{Wj_1} = m_t \pm 50\,$GeV, $p_{T^{j_2}}>800\,$GeV and the relevant KKG mass window $2.5
\; \hbox{TeV} < m_{Wjj}<3.5\,$TeV is 25fb. 
Applying a $b$ mistag probability of $3$\% (see~\cite{KKgttbar}) 
and leptonic BR of $2/9$ for $W$ further reduced this cross section to about 0.2 fb.
We compare this to our top pair production cross section
(signal+background) satisfying these cuts 
of 80 fb which is reduced to about $5$ fb after applying $b$-tagging
and including BR's (see details below). 
Thus, we conclude the $Wjj$ background to be small.
We found that $Wb\bar{b}$ and single top production with these same cuts have even smaller
cross section than $Wjj$ after including BR and $b$-tagging efficiency.
The results of our particle level analysis can be seen in Fig.~\ref{kkg_4b_tt}.

Following the procedure in ~\cite{ttbarspin}, the neutrino is reconstructed using a zero transverse momentum hypothesis on the event, 
with the neutrino carrying away the missing momentum. 
We required the lepton and neutrino to reconstruct an on-mass-shell $W$ ($M_{ {\it l} \nu } = M_W = 81\,GeV$).
This information is sufficient to reconstruct the neutrino momentum, modulo a quadratic ambiguity. 
In the case where we obtain two solutions, we used the one which better reconstructs the top ($|M_{W}^{lep}-M_{top}|<50\,$GeV). 
Additional studies, beyond the scope of this work, are required to characterize the effects of $W$ reconstruction 
when the lepton and neutrino are nearly collinear at high energies. 
We address this issue by noting that in our data sample, 
we were able to impose a $\Delta R = 0.15$ separation 
between the lepton and neutrino with minimal loss of statistics.
The cuts and other kinematical constraints are summarized in Table~\ref{tab:sel_cuts}.
  
In the following sections, we present our results from both partonic- and particle-level analyses.
We shall see that these two analyses are consistent with each other, 
and that no significant bias was introduced due to our selection cuts or reconstruction procedures.
\footnote{The lepton $P_{T}$ and $\mbl$ cuts are particularly scrutinized. Their impact on phase space will directly affect our polarization analysis.}
We remind the reader that we did not perform detailed detector simulation and hence, have not included the resulting smearing effects. 
We expect that, due to the nature of our kinematical region, 
the dominant smearing will be of ${\cal O}(3\%)$ (see {\it e.g.}~\cite{TDR}) which will induce small corrections to our mass resolution. 
A study of how the detector effects will modify the polarization asymmetry (discussed below) is beyond the scope of this work and will discussed in~\cite{KPV}.
  
{\small
\begin{table}[htbp]
\begin{center}\hspace*{.1cm}
\begin{tabular}{|l||c|c|c|}
\hline
   {\bf Selection}  &  {\bf Variables}               & {\bf  Cuts}      \\
\hline
\hline\hline
                    &   lepton                 &  $p_T > 10 \,{\rm GeV},\,|\eta|<2.5$ \\ 
  Kinematic    &  $\ge$ 2 jets            &  $p_T > 30 \,{\rm GeV},\,|\eta|<2.5$ \\
    and     &  tagged $b$-jets         &  $\ge$ 1\\
  acceptance                    &  missing energy ($\nu$)  &  $p_T^{miss} 
> $ 20 GeV        \\
                    &  lepton isolation &  $\Delta R\geq$ 0.4 (non-$b$-jets)         \\                          
                    &  $b$-jet lepton isolation  &  $\Delta R\geq$ 0.4 {\em or} \\
 & & $m_{bl}
\geq$ 40 GeV        \\ 
                               
\hline \hline
 Reconstruction     &  $|M_W^{\rm had}-M_W|$ &  $ < $ 50 GeV  \\
    quality         &  $|M_t^{\rm had}-M_t|$ &  $ < $ 50 GeV  \\
     for \#jets$>2$,&  $|M_t^{\rm lep}-M_t|$ &  $ < $ 50 GeV  \\
 2 $b$-jets required   &        &    \\
\hline
 Reconstruction     &    &    \\
    quality         &  $|M_t^{\rm lep}-M_t|$ &  $ < $ 50 GeV  \\
    a $b$-jet+$t$-jet&    ``top-jet'' &    $p_{T^{t}}>$ 800 GeV     \\
 \hline
\end{tabular}
\vspace*{.2cm}
\caption{\it Selection cuts in the semileptonic $t\bar{t}$ channel.} 
\label{tab:sel_cuts}
\end{center}
\end{table}
}

\subsubsection{Differential cross section}
\label{diffcs}

%
%
The SM top pair production rate falls steeply as a function of the invariant mass. 
The uncertainty from PDF's in this {\em shape} is far less than that in the total cross-section. 
Hence we look for a signal from KK gluons in the {\em differential} $t\bar{t}$ cross-section as opposed to simply counting the total number of $t \bar{t}$ events. 
We do not expect a sharp resonance in this distribution due to the large width of the KK gluon, 
but we do obtain a statistically significant ``bump'' as discussed below.

\begin{figure}[htbp]
\includegraphics[width=0.48\textwidth]{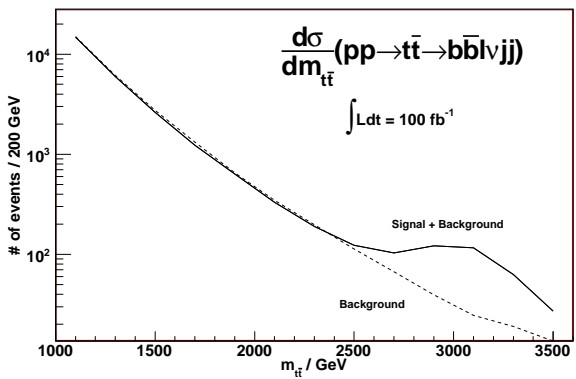}
\caption{\label{kkg_4a_tt}
Invariant $t\bar{t}$ mass distribution for $\mkkg = 3\,$TeV production at the LHC. 
The solid curve presents signal+background distribution, while the dashed curve presents the $\ttbar$ SM background, based on partonic level analysis.}
\end{figure}

\begin{figure}[htbp]
\includegraphics[width=0.48\textwidth]{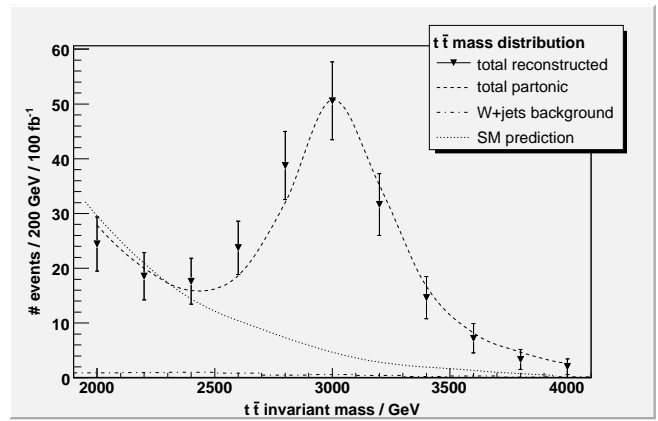}
\caption{\label{kkg_4b_tt}
Invariant $t\bar{t}$ mass distribution for 3 TeV KKG, focusing on the area near the peak. 
The error bars correspond to statistical uncertainties and represent our particle level analysis. 
The dotted line stands for the SM prediction. 
The dashed-dotted line shows the $Wjj$ background. 
The dashed line shows the signal+background from Sherpa's partonic level analysis.
%
%
}
\end{figure}

The differential cross section as a function of $m_{t\bar{t}}$ is shown in 
Figs.~\ref{kkg_4a_tt}
and \ref{kkg_4b_tt} for $\mkkg = 3\,$TeV produced at the LHC. 
In Fig.~\ref{kkg_4a_tt} we compare the total (signal + background) distribution to the SM (background) distribution, based on a partonic-level analysis.
In Fig.~\ref{kkg_4b_tt}, we focus on the area near the peak and 
we consider contributions from the reducible background (from $Wjj$).
We show the particle level results and 
the corresponding statistical uncertainties of event reconstruction.
%
%
The predictions for the SM and SM+RS models, based on 
partonic-level analysis
(same as in Fig.~\ref{kkg_4a_tt}), are also shown 
for comparison.
We see that, since the partonic and particle level data are consistent 
with each other, we do not expect a large bias in the ability to 
reconstruct the KKG mass. 

In the following we describe the
reconstruction efficiency and how we estimate our signal to background
ratio and the sensitivity to the KK gluon mass based on this analysis.
Following~\cite{KKgttbar}, we assume a 20\% efficiency for tagging $b$-jets
($\epsilon_b$), independent of the $b$-jet energy. Our particle level study
shows that the efficiency of the additional cuts described, $\epsilon_{\rm
cut}$, in Table~\ref{tab:sel_cuts} for the reconstruction of $t\bar{t}$
system in the mass window around KKG
is about 20(21)\% for $\mttbar = 3(4) $TeV. We
find that for the SM the reconstruction efficiency is lower,   9(10)\% for
$\mttbar = 3(4)$ TeV.
The signal+background (BG+KKG) and background (BG) reconstruction efficiencies 
differ
because
the 
BG and BG+KKG events have different kinematics.
The background is
dominated by $gg$ fusion events which are more forwardly-peaked
in the top pair center of mass (cm) frame than the $q \bar{q}$ fusion 
events. 
%
%
Hence, the $gg$ events have a 
smaller $P_T$\footnote{Note that, inside the mass window, 
%
%
the total momentum/energy of each top quark
in cm frame is roughly fixed at $M_{ KK G} / 2$.}
than the $q \bar{q}$ events.
Since KK gluon signal comes only from $q \bar{q}$ fusion, the 
$p_T$ cut on the top-quark
reduces background more than the signal.

In addition, the branching ratio for the $lj$ decay is given by 
$BR_{lj}=2 \times 2/9 \times 2/3 \simeq 0.3$. 
The total efficiency is given by $BR_{lj}\times \epsilon_{\rm cut}\times \epsilon_b \sim 1\%$.

We estimate the statistical significance of our signal by looking at the bump.
An invariant $t\bar{t}$ mass window cut $0.85 M_{KKG} < M_{t\bar{t}} < 1.5 M_{KKG}$ is applied.
The lower bound corresponds roughly to the width. The upper bound is not particularly important due to the steep falloff in cross section.
Below the $\mkkg$ threshold, the signal+background distribution is actually below the background one due to destructive interference. 
Therefore, we choose an asymmetric mass window cut.
We estimate the ratio of the signal, $S$, to the statistical error in the the background, $\sqrt{B}$, via our particle level analysis in the mass window, 
for 100 fb$^{-1}$
We find
\begin{eqnarray}
S/\sqrt B &\approx& 11.0 \qquad {\rm for}\ \ \ \  M_{KKG}=3\,{\rm TeV}\,,\nonumber\\
S/\sqrt B &\approx& 4.2 \qquad {\rm for}\ \ \ \ M_{KKG}=4\,{\rm TeV}
\end{eqnarray}
In addition we find the following values for signal over background:
\begin{eqnarray}
S/B &\approx& 2.0\qquad {\rm for}\ \ \ \  M_{KKG}=3\,{\rm TeV}\,,\nonumber\\
S/B &\approx& 1.6\qquad {\rm for}\ \ \ \ M_{KKG}=4\,{\rm TeV}
\end{eqnarray}
where the total number of events inside the mass window for $\mkkg = 4\,$TeV 
that pass all cuts is ${\cal O}(10)$. 
Thus for 100fb$^{-1}$ we estimate the LHC reach to be below 4 TeV for the KK gluon mass.
We discuss below the use of discriminators which may improve this analysis.
One should stress
that Figs.~\ref{kkg_4a_tt} and \ref{kkg_4b_tt}
demonstrate a clear evidence for
a bump  in the differential cross-section.
Such a deviation in  shape from the background
distribution  as well as good $S/B \simeq 2$
ratio guarantee that the KKG signal will be
clearly seen for $M_{KKG}$ below about 4 TeV.
A more sophisticated analysis 
could possibly improve further the significance 
and signal-to-background ratio.

\subsubsection{Polarization asymmetry}

We now consider how measurement of the polarization of the ultra
relativistic top pairs provides us with an important tool for detection 
of the KK gluon.
The fact that the KK gluon decays mostly into two tops turns out to be
advantageous because the top quark decays before it hadronizes. 
Therefore, the top spin/chirality information is encoded in the distribution of its decay products. Moreover,
since we are dealing with very energetic top quarks, their masses can be neglected and their chirality is conserved.
The SM top pair production is dominated by parity invariant QCD processes, 
so we expect to generate an (almost) equal number of left- and right-handed pairs.
However, in the RS1 model that we are considering, 
we expect a strong bias towards RH tops (from KK gluon decays) so a {\em large} left-right (LR) asymmetry is expected.

We can include EW production processes in the SM.
Note that in the ultra-relativistic case we can neglect effects related to EWSB. In this case, the SM EW production processes can be characterized by the hyper-charge and 
weak coupling separately. The latter is stronger and couples only to LH particles~\cite{KPV}. 
Thus we get a sharp prediction that the deviation of $\PLR$ from zero
in the SM (due to EW processes)
carries the opposite sign compared to the above RS1 KK gluon signal 
(again, in the latter, the RH top dominates). 
The EW processes can only be mediated via $\qqbar$ annihiliation processes. 
To summarize, the SM $\PLR$ is suppressed by $g_2^2 / g_{ QCD }^2\sim 0.35$ and the ratio between the $\qqbar$ and $gg$ production rates\footnote{this is 
probably the only source of uncertainty for the value of this asymmety.}
and, hence is much smaller than the $O(1)$ asymmetry expected in the RS1 model from KK gluon decays, in addition to having the opposite sign to the RS1 signal.

The RS1 prediction can be tested via measurement of $\PLR$ of the top pairs sample as follows.
The angular distribution of the positron from a purely RH and LH top
quark decay is given by\cite{ttbarspin,ttbarspinmore}:
\begin{eqnarray}
\frac{ d N }{ d \cos \theta } & \sim& ( 1 \pm \cos \theta )
\end{eqnarray}
where $\theta$ is angle between the positron direction in the {\em rest}
frame of the top and the direction of the top quark boost (in the
parton/$\ttbar$ center of mass frame).\footnote{the latter
is also the top spin quantization axis.} It is useful to define the {\em
polarization asymmetry} via a ``forward-backward'' asymmetry as
\begin{eqnarray}
\PLR & \equiv & 2 \times 
\frac{ N_+ - N_-}{ N_+ + N_-},
\end{eqnarray}
where $N_+ \equiv \int_{ 0 }^{ \pi / 2 } d \cos \theta d N /d \cos \theta$ is the number of positrons emitted (in the rest frame of the top) along the direction of the top quark boost 
(and similarly for $N_-$). 
For purely RH (LH) top quark, we get $\PLR = \pm 1$.

We used Sherpa, which supports spin/helicity amplitudes, to numerically analyze the signal and background. 
As mentioned above, the asymmetry is measured relative to the direction of the top quark boost in the center of mass frame of the top pair. 
The challenge here is to reconstruct the top rest frame from observables in the event.
%
%
The lepton is boosted into the cm frame, and subsequently reboosted 
into the top quark rest frame, using essentially the $P_T$ of the top quark.
%
%

\begin{figure}[htbp]
\includegraphics[width=0.48\textwidth]{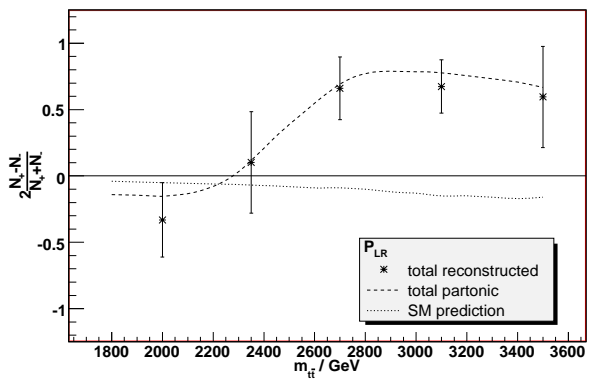}
\caption{\label{PLR}
$\PLR(\mttbar)$ for $\mkkg = 3\,$TeV: The error bars correspond to statistical uncertainties and represent particle level analysis.
The dotted line stands for the SM prediction. The dashed line shows the signal+background from Sherpa's partonic level analysis.
%
%
}
\end{figure}

The LR polarization asymmetry as a function of $m_{t\bar{t}}$ is shown
in Fig.~\ref{PLR} for $M_{KKG}=3\,$TeV with 100$\,$fb$^{-1}$ data.
 The error bars correspond to statistical uncertainties and represent
  our particle level analysis using Sherpa. 
%
%
We also show the
  signal+background from partonic level analysis using Sherpa.  
%
%

Note that the leptons from $t_R$ tend to be emitted in the forward direction, whereas the opposite is true for leptons from $t_L$.
Therefore, the $P_T$ cut of the lepton will non-trivially impact the asymmetry, due to the kinematics and small masses of the leptons.
We chose a $P_T$ cut of 10 GeV, whose effects were manageable 
as we checked
via Monte Carlo simulations.

We see that the partonic and particle level data are consistent with each other. 
Therefore, we have not introduced any significant bias in the observed asymmetry as a result of the above cuts and reconstruction procedure.
We remind the reader that we do not characterize herein the detector
effects on the $\PLR$, which will add to the uncertainty in
reconstruction of the $\ttbar$ cms and top rest frame.

Note that for $\mttbar\ll M_{KKG}$ the asymmetry is negative and close to zero (for both curves) as expected since the SM production is dominant here~\cite{KPV}. 
On the other hand, for $\mttbar \sim M_{KKG}$, a sizable asymmetry is obtained for the 
signal+background curve with a {\em positive} sign which implies a 
significant excess of RH $\ttbar$ 
as expected in the RS1 model~\cite{custodial, aps}. 
Correlated observations of
such a sizable asymmetry and an excess in
the differential cross-section for the {\em same} $\mttbar$ (as in 
Figs.~\ref{kkg_4a_tt} and \ref{kkg_4b_tt})
will be a strong evidence for a KK gluon.
Also, the asymmetry for SM background increases (but still remains small) 
with $\mttbar$ 
since the ratio of $q \bar{q}$ fusion (which gives the asymmetry) to
$gg$ fusion (which is symmetric) increases with higher $\mttbar$.

As already mentioned, we are in the relativistic limit for the tops produced from the KK gluon so that the spins of the top pair are correlated,
independent of RH or LH dominating in the KK gluon decay (this holds for any chiral theory). 
Therefore, 
an 
analysis similar to that
for the lepton from $t$ decay (mentioned above)
can be applied for the $b$ and light jets 
emitted from $\bar t$ decay on the other side
(although these decay products are not as powerful spin analysers as the lepton). 
This would further increase the statistics and the significance of our signal. 
The required analysis is more involved and beyond the scope of this work.
However, we expect that such an analysis, even though less precise, 
may allow us to eliminate some of the uncertainties due to biases and other systematic effects.

\subsubsection{Signal versus background optimization}
\label{optimize}

As we already indicated in section \ref{diffcs}, the
$P_T$ cut reduces background more as compared to signal.
In this section, we discuss possible
{\em additional} 
cuts which can be applied to the analyses of the differential 
cross-section and $P_{ LR }$ to improve the significance of our results.

A cut on the forwardness of the $\ttbar$ pairs 
in cm frame 
is useful for removal of the background. 
The reason is that (as already mentioned in section \ref{diffcs}) 
the top quarks produced from gluon fusion (via top $t$-channel exchange) 
tend to be more forwardly-peaked than the ones produced from $q\bar{q}$ annihilation.
KK gluons are produced only through $q \bar{q}$ annihilation, 
whereas the SM background is dominated by $gg$ fusion. 
Therefore, an appropriate cut on $\eta^*$ 
(rapidity in cm frame)
of each top quark will eliminate a substantially larger part of the SM QCD background, 
at the expense of a smaller fraction of the signal.

We applied the cut $|\eta|^*<1.8$ to find that
it has virtually no effect on the signal (as desired and as expected).
%
%
Whereas, 
we find that
the SM background reduced
(and hence our significance increased) by only 
$O( 10 \% )$, perhaps unlike the
expectation of a more significant reduction in background.
The reason is that we find the $P_T$ cut on hadronic top (the top jet),
which is part of our event selection cuts, and the $|\eta|^*<1.8$ cut to be 
correlated (as expected from the discussion in section \ref{diffcs}).

Note that the only
reason we included this $P_T$ cut as part of our event selection was
that we could not reconstruct the hadronic top in the conventional manner.
With a more sophisticated analysis for reconstruction
of the hadronic top (for example, 
resolving sub-structure in top-jet as mentioned before),
this $P_T$ cut might {\em not} be required as part of event selection.
In the absence of $P_T$ cut, the
$|\eta^*| <1.8$ cut might then reduce background more significantly.
However, given our $P_T$ cut, the only 
possibility for the $|\eta^*|$ cut to be useful in removing background
seems to be to cut on smaller values of $|\eta^*|$. 
Due to limited statistics for $M_{KKG}\sim3\,$TeV and $100\,$fb$^{-1}$,
we might not be able to apply such a stronger cut.
However, we note that
such a cut can be applied in case of higher luminosity or a lower KK mass.
We leave a more detailed study for the future.
%
%
%
%

Next, we apply the $| \eta^* | <1.8$ cut to the analysis of polarization asymmetry.
%
%
We find that
%
%
the (negative) asymmetry for the SM background increases 
by $O( 10 \% )$
after applying 
this cut. 
The reason is that these cuts increase the fraction of $q \bar{q}$ 
fusion events (compared to $gg$ fusion) in the sample 
-- again, only $q \bar{q}$ fusion contributes to the asymmetry.
%
%
%
%
Furthermore we find that signal+background 
$P_{ LR }$ is 
not significantly affected (within
the statistical errors). 
Again,
the reason for only a small effect of $| \eta^* |$
cut is the correlation between the 
$| \eta^* | <1.8$  and $P_T$ cuts (as mentioned
above in the case of differential cross-section).

Finally we want to comment about the possibility of using a cut on the boost to distinguish signal vs. background 
which may be useful for lower KK masses as follows.
The gluonic content of the protons is symmetric between the two incoming protons, 
implying that the $\ttbar$ pairs from the $gg$ fusion production will be mostly produced with a small boost.
The $\qqbar$ annihilation production, however, proceeds through the asymmetric $q \bar{q}$ content of the proton~\cite{Tau}.  
Thus, we expect the corresponding top pairs to exhibit a larger boost.  
This in principle implies that by applying suitable cuts on this boost 
($\beta_{\rm cm}$) one can purify the (signal) sample,
obtaining a larger polarization asymmetry and a larger significance from the differential cross-section analysis.
However,
using a partonic level study, we find that this cut 
is effective only for rather low KK gluon masses (at or below the $1$ TeV scale).

\section{Electroweak sector \& alternative quark configuration}
 
We now discuss briefly the electroweak (EW) gauge KK modes. 
As mentioned before, the cross-section for KK $Z/W$ production (via $q \bar{q}$ fusion) is smaller than that of the KK gluon by $\sim g_Z^2 / g_{ QCD }^2$.
As for the KK gluon, fermionic decays of KK $Z$ are dominated by top quarks. 
Its leptonic decays are highly suppressed.
Thus, the KK $Z$ also contributes to the excess $t \bar{t}$, but is subdominant to the KK gluon signal in this channel.

However, there is a new feature in the excess $t \bar{t}$ sample due to the KK $Z$ contribution. 
The couplings of the KK $Z$ to the {\em initial} state are non vector-like, unlike in the case of the KK gluon. 
Combined with non vector-like couplings to the final state top quarks, 
we obtain the usual forward/backward asymmetry $A_{ FB }$ at the level of $\sim g_Z^2 / g_{ QCD }^2$ in the excess top pair sample. 
Note that an asymmetry of this size is present even in the SM due to $Z$ exchange. 
The crucial point is that sign of this asymmetry is different than in the SM since $t_R$ dominates in the final state as opposed to $t_L$ in the SM.
We can measure this asymmetry (both in the SM and in RS1 model) since we know the direction of $q$ or forward (vs. $\bar{q}$ or backward)
based on the direction of the boost \cite{KPV}.

As mentioned before, the KK $W/Z$ also have sizable decays to Higgs, including longitudinal $W/Z$. 
As a corollary, {\em production} of KK $Z/W$ via longitudinal $W/Z$ fusion can be important. We plan to study such signals in the future.

\subsection{Effects of enhanced $b_L$ coupling to KK gluon}\label{SecZbb}

As indicated above, the $b_L$ coupling to the KK gluon is larger ($\sim g_{ QCD }$) than to light quarks (including $b_R$).  
In fact, with the symmetry protection for $Z b_L b_L$ coupling \cite{Zbb}, 
$(t,b)_L$ can be localized very close to  the TeV brane so that the $b_L$ coupling to the KK gluon can be as large as $\xi g_{ QCD }$. 
Hence,  $b_L \bar{b}_L$ fusion might become the dominant production mechanism for KK gluon.

Since both $b$ and $\bar{b}$ are sea partons and have the same content
inside a proton,  the excess top events from $b_L \bar{b}_L$ fusion into
KK gluon are less boosted events, but are also less forward than from
$gg$ fusion.  Recall that the excess from $q \bar{q}$ fusion is more
boosted and less forward than $gg$ fusion. Hence, the $\eta^{*}$ cut might
still be useful, as before, to enhance the signal over background,  but the
$\beta_{ \rm cm }$ cut might be less useful in enhancing the signal.

A new feature from $b_L \bar{b_L}$ fusion into KK gluon is that, due to
vector-like couplings in both initial (cf. coupling to light quarks) and
final states, it will result in a $A_{ FB }$ in KK {\em gluon} top
events (cf. $A_{ FB }$ in $q \bar{q}$ fusion events is only from SM or
KK $Z$). However, we cannot measure this $A_{ FB }$ since  we do not
know forward ($b$) vs. backward ($\bar{b}$) direction due to absence of
sizable boost (cf. in $q \bar{q}$ fusion). 

The excess top events from $b_L \bar{b}_L$ fusion into KK gluon will
have the same non-zero $P_{ LR }$ as the excess from $q \bar{q}$ fusion
(again, the excess from $b_L$ fusion will be in less boosted  events
compared to that from $q \bar{q}$ fusion into KK gluon).  In fact, in
the extreme case of $(t,b)_L$ being very close to the TeV brane and
$t_R$ having close to a flat profile, we see that the sign of $P_{ LR }$
in signal will be reversed compared to what we discussed before (i.e.,
will be $< 0$).  This sign is same as in the SM, but the crucial point
is that the $O(1)$ size is much larger than that expected in the SM.

\section{Conclusions}
In summary, the framework of a warped extra dimension provides a novel and very interesting resolution to the Planck-weak {\em and} flavor hierarchy problem of the SM.
It tends to generically single out the top quark with enhanced couplings to the new states, 
whereas couplings to light fermions, in particular to proton's constituents, are suppressed. 
These features make it challenging to detect the new states.

In spite of this challenge, we have shown that the production of the KK gluon with subsequent decays to top pairs at the LHC is a very interesting channel, 
which would be worthwhile to explore further. 
In particular, for $100$ fb$^{-1}$ integrated luminosity, we demonstrated that one can discover the KK gluon with the mass 
$M_{KKG} 
\lesssim $ 4 TeV 
based on the {\em correlated} observations 
of an excess in the top pair differential cross-section  and a sizable left-right polarization asymmetry ($P_{ LR }$). 
This asymmetry is much larger than in the SM due to very different couplings of the KK gluon to RH and LH top quarks. 
We discussed how 
a cut on transverse momentum of top quarks reduces background
compared to the signal and how 
it might be possible to
further improve the signal to background ratio by imposing cuts on 
the boost of the top center of mass frame in the laboratory frame
and forwardness of top pairs in the parton center of mass frame. 
We briefly discussed the EW sector which requires more study. 
Its detection is similarly challenging due to suppressed couplings to the proton's constituents  
--
in fact, it has lower production rate than for the KK gluon -- and suppression of decays to leptons (``golden'' decay modes). 

Finally, we emphasize that, via the AdS/CFT duality \cite{Maldacena:1997re}, the RS framework should be viewed as a tool to study $4D$ strong dynamics. 
%
%
In fact, the idea of a composite, pseudo-Goldstone boson (PGB), Higgs in $4D$ has been studied in the RS framework (called ``holographic'' PGB Higgs) \cite{Contino:2003ve}
It is therefore likely that our results apply (in general) to $4D$ TeV-scale strong dynamics responsible for EWSB. 
In particular, our analysis with regards to the LHC signals leads to the following observation about other frameworks 
which address the little hierarchy problem and rely on UV completion via strong dynamics ({\it i.e.}, little Higgs and some flat extra dimensional models).  
According to the belief that the RS1 framework can be used to obtain intuition about such models
\footnote{In fact, see reference \cite{Thaler:2005en} for UV completion of the 
Litte{\em st} Higgs model using RS framework.},
our studies suggest that  these models might be characterized by LHC signals which are somewhat different from those usually emphasized in the literature. 
The reason is that the couplings between the extended electroweak sector and the 
light (heavy) {\em SM} particles may be actually highly suppressed (enhanced), 
unlike what is typically assumed in other LHC studies.
\footnote{References \cite{Perelstein:2003wd} do mention, in the context of LHC signals, 
that suppressed couplings of light fermions to $Z^{ \prime }$, $W^{ \prime }$ are motivated in order to satisfy electroweak precison tests. 
However, most of these studies still assume {\em universal} fermionic couplings so that couplings to top quark are also suppressed in this case.
Whereas, we emphasize that top quark couplings to the new states are likely to be enhanced, leading to difficulties in detection of new states.}
Generically, the new particles will be broad,  with small production rates and non-leptonic decay channels. 
As such, these models may face similar challenges as that for the KK gluon, in the detection of new states.

\mysection{Acknowledgements}
K.~A.~ and G.~P.~ thank the Aspen Center for Physics for their hospitality.  We thank
Marco Battaglia, Tao Han, Beate Heinemann, Ian Hinchliffe, Ayana Holloway,  Hitoshi Murayama, Frank Paige, Michele Papucci, Frank
Petriello, Marjorie Shapiro and George Sterman  for discussions. 
We thank the Sherpa team, especially Tanju Gleisberg, for implementing
an RS1 model in the MC event generator.
The work of A.B. was supported by the US National Science Foundation under award PHY-0555545.
The work of T.K. was supported by DOE Grant No. DE-AC02-98CH10886.

\end{document}